%% file: prl2.tex
\documentclass[twocolumn,showpacs,aps,prl,amsmath,amssymb,
superscriptaddress,letterpaper]{revtex4}

\usepackage{graphicx}
\usepackage{dcolumn}
\usepackage{bm}

\input def.sty

\begin{document}

\title{Precision Measurement of the Weak Mixing Angle in 
  \Moller\ Scattering} 

\input authors.tex

\date{April 26, 2005}

\begin{abstract}
We report on a precision measurement of the parity-violating asymmetry
in fixed target electron-electron (\Moller) scattering: 
$A_{PV} = (-131\pm 14\ \stat\ \pm 10\ \syst)
\times 10^{-9}$, leading to the determination of the weak
mixing angle 
$\stwrun = 0.2397\pm 0.0010\ \stat\ \pm 0.0008\ \syst$,
evaluated at $Q^2 = 0.026~\mathrm{GeV}^2$. 
Combining this result with the measurements of \stwrun at the $Z^0$ pole, 
the running of the weak mixing angle is observed with over
$6\sigma$ significance. The measurement sets constraints on
new physics effects at the \tev scale. 
\end{abstract}

\maketitle

\input intro

\input expt

\input analysis

\input results

\input conclusion

\input ack

\input bib
\end{document}

%% file: authors.tex
\affiliation{University of California, Berkeley, California 94720}

\affiliation{California Institute of Technology, Pasadena,
California 91125}
\affiliation{University of Massachusetts, Amherst, Massachusetts 01003}

\affiliation{Princeton University, Princeton, New Jersey 08544}

\affiliation{CEA Saclay, DAPNIA/SPhN, F-91191 Gif-sur-Yvette, France}

\affiliation{Smith College, Northampton, Massachusetts 01063}

\affiliation{Stanford Linear Accelerator Center, Menlo Park,
California 94025}

\affiliation{Syracuse University, Syracuse, New York  13244}

\affiliation{Thomas Jefferson Laboratory, Newport News, Virginia
23606}

\affiliation{University of Virginia, Charlottsville, Virginia
22903}

\author{P.L.~Anthony}\affiliation{Stanford Linear Accelerator Center, Menlo Park, California 94025}
\author{R.G.~Arnold}\affiliation{University of Massachusetts, Amherst, Massachusetts 01003}
\author{C.~Arroyo}\affiliation{University of Massachusetts, Amherst, Massachusetts 01003}
\author{K.~Bega}\affiliation{California Institute of Technology, Pasadena, California 91125}
\author{J.~Biesiada}
\affiliation{University of California, Berkeley, California 94720}
\affiliation{Princeton University, Princeton, New Jersey 08544}
\author{P.E.~Bosted}
\affiliation{University of Massachusetts, Amherst, Massachusetts 01003}
\affiliation{Thomas Jefferson Laboratory, Newport News, Virginia 23606}
\author{G.~Bower}\affiliation{Stanford Linear Accelerator Center, Menlo Park, California 94025}
\author{J.~Cahoon}\affiliation{University of Massachusetts, Amherst, Massachusetts 01003}
\author{R.~Carr}\affiliation{California Institute of Technology, Pasadena, California 91125}
\author{G.D.~Cates}\affiliation{University of Virginia, Charlottsville, Virginia 22903}
\author{J-P.~Chen}\affiliation{Thomas Jefferson Laboratory, Newport News, Virginia 23606}
\author{E.~Chudakov}\affiliation{Thomas Jefferson Laboratory, Newport News, Virginia 23606}
\author{M.~Cooke}\affiliation{University of California, Berkeley, California 94720}
\author{P.~Decowski}\affiliation{Smith College, Northampton, Massachusetts 01063}
\author{A.~Deur}\affiliation{University of Virginia, Charlottsville, Virginia 22903}
\author{W.~Emam}\affiliation{Syracuse University, Syracuse, New York  13244}
\author{R.~Erickson}\affiliation{Stanford Linear Accelerator Center, Menlo Park, California 94025}
\author{T.~Fieguth}\affiliation{Stanford Linear Accelerator Center, Menlo Park, California 94025}
\author{C.~Field}\affiliation{Stanford Linear Accelerator Center, Menlo Park, California 94025}
\author{J.~Gao}\affiliation{California Institute of Technology, Pasadena, California 91125}
\author{M.~Gary}\affiliation{University of California, Berkeley, California 94720}
\author{K.~Gustafsson}
\altaffiliation[Now at ]{Helsinki Institute of Physics, Finland.}
\affiliation{California Institute of Technology, Pasadena, California 91125}
\author{R.S.~Hicks}\affiliation{University of Massachusetts, Amherst, Massachusetts 01003}
\author{R.~Holmes}\affiliation{Syracuse University, Syracuse, New York  13244}
\author{E.W.~Hughes}\affiliation{California Institute of Technology, Pasadena, California 91125}
\author{T.B.~Humensky}\affiliation{Princeton University, Princeton, New Jersey 08544}
\author{G.M.~Jones}\affiliation{California Institute of Technology, Pasadena, California 91125}
\author{L.J.~Kaufman}\affiliation{University of Massachusetts, Amherst, Massachusetts 01003}
\author{L.~Keller}\affiliation{Stanford Linear Accelerator Center, Menlo Park, California 94025}
\author{Yu.G.~Kolomensky}\affiliation{University of California, Berkeley, California 94720}
\author{K.S.~Kumar}
\affiliation{University of Massachusetts, Amherst, Massachusetts 01003}
\author{P.~LaViolette}\affiliation{University of Massachusetts, Amherst, Massachusetts 01003}
\author{D.~Lhuillier}\affiliation{CEA Saclay, DAPNIA/SPhN, F-91191 Gif-sur-Yvette, France}
\author{R.M.~Lombard-Nelsen}\affiliation{CEA Saclay, DAPNIA/SPhN, F-91191 Gif-sur-Yvette, France}
\author{Z.~Marshall}\affiliation{University of California, Berkeley, California 94720}
\author{P.~Mastromarino}\affiliation{California Institute of Technology, Pasadena, California 91125}
\author{R.D.~McKeown}\affiliation{California Institute of Technology, Pasadena, California 91125}
\author{R.~Michaels}\affiliation{Thomas Jefferson Laboratory, Newport News, Virginia 23606}
\author{J.~Niedziela}\affiliation{University of Massachusetts, Amherst, Massachusetts 01003}
\author{M.~Olson}\affiliation{Stanford Linear Accelerator Center, Menlo Park, California 94025}
\author{K.D.~Paschke}\affiliation{University of Massachusetts, Amherst, Massachusetts 01003}
\author{G.A.~Peterson}\affiliation{University of Massachusetts, Amherst, Massachusetts 01003}
\author{R.~Pitthan}\affiliation{Stanford Linear Accelerator Center, Menlo Park, California 94025}
\author{D.~Relyea}\affiliation{Princeton University, Princeton, New Jersey 08544}
\affiliation{Stanford Linear Accelerator Center, Menlo Park, California 94025}
\author{S.E.~Rock}\affiliation{University of Massachusetts, Amherst, Massachusetts 01003}
\author{O.~Saxton}\affiliation{Stanford Linear Accelerator Center, Menlo Park, California 94025}
\author{J.~Singh}\affiliation{University of Virginia, Charlottsville, Virginia 22903}
\author{P.A.~Souder}\affiliation{Syracuse University, Syracuse, New York  13244}
\author{Z.M.~Szalata}\affiliation{Stanford Linear Accelerator Center, Menlo Park, California 94025}
\author{J.~Turner}\affiliation{Stanford Linear Accelerator Center, Menlo Park, California 94025}
\author{B.~Tweedie}\affiliation{University of California, Berkeley, California 94720}
\author{A.~Vacheret}\affiliation{CEA Saclay, DAPNIA/SPhN, F-91191 Gif-sur-Yvette, France}
\author{D.~Walz}\affiliation{Stanford Linear Accelerator Center, Menlo Park, California 94025}
\author{T.~Weber}\affiliation{Stanford Linear Accelerator Center, Menlo Park, California 94025}
\author{J.~Weisend}\affiliation{Stanford Linear Accelerator Center, Menlo Park, California 94025}
\author{M.~Woods}\affiliation{Stanford Linear Accelerator Center, Menlo Park, California 94025}
\author{I.~Younus}\affiliation{Syracuse University, Syracuse, New York  13244}

\collaboration{SLAC E158 Collaboration}\noaffiliation

%% file: intro.tex

Precision measurements of weak neutral current processes at low
energies rigorously test the Standard Model of electroweak
interactions. 
Such measurements are sensitive to new physics effects
at \tev energies, and are complementary to searches 
at high energy colliders.

One class of low-energy electroweak measurements involves 
scattering of longitudinally polarized electrons from unpolarized
targets, allowing for the determination of a parity-violating asymmetry
$\APV\equiv{(\sigma_R-\sigma_L)}/{(\sigma_R+\sigma_L)}$, where
$\sigma_{R(L)}$ is the cross section for incident right(left)-handed
electrons.  \APV arises from the interference of the weak and
electromagnetic amplitudes \cite{zeld} and is sensitive to the electroweak
coupling constants and thus the weak mixing angle
$\theta_{\mathrm{W}}$.

The electroweak parameter \stwrun\ is defined as the ratio of the
electromagnetic to the weak isospin coupling constants
\cite{ref:SM}. Possible new physics contributions at very high energy
scales can be expressed in terms of their impact on the
measured value of \stwrun. Measurements at low momentum transfers 
$Q^2\ll M_Z^2$ can have sensitivity comparable to high  
energy collider searches for new physics provided \stwrun\ is measured
to better than 1\%.  

At such a precision, the variation of the coupling constants with momentum
transfer, a fundamental property of gauge interactions referred to 
as running \cite{ref:running}, must be taken into account. While the
running of the electromagnetic and strong coupling constants has been clearly
established, it has not been unambiguously observed for \stwrun so
far. The variation 
of \stwrun\ from $Q^2\approx 0$ to $Q^2= M_Z^2$ is due to
higher order 
amplitudes involving virtual weak vector bosons and fermions in quantum loops, 
referred to as electroweak radiative
corrections~\cite{marciano1,otherad}.

To date, the most precise low-energy determinations of the weak mixing angle 
come from studies of parity violation in atomic transitions
\cite{ref:APVcs} and measurements of the neutral current to charge
current cross section ratios in 
neutrino-nucleon deep inelastic scattering \cite{ref:nutev}. 
In this Letter, we present a measurement of the weak mixing angle in
electron-electron (\Moller) 
scattering, a purely leptonic reaction with little theoretical uncertainty. 
We have previously reported the first
observation of \APV in \Moller\ scattering \cite{E158prl}. 
Here, we report on a significantly improved measurement of \APV
resulting in a precision determination of \stwrun at low momentum
transfer.

%% file: expt.tex
%
%

At a beam energy of $\simeq 50$ GeV available at End Station A at SLAC
and a center-of-mass scattering angle of 
$90^\circ$, 
$A_{PV}$ in \Moller\ scattering is predicted to be $\simeq 320$ 
parts per billion (ppb) at tree level \cite{darman}. Electroweak 
radiative corrections 
\cite{marciano1,otherad}
and the experimental acceptance reduce the measured asymmetry 
by more than 50\%.

The principal components of the experimental apparatus 
are the polarized electron beam, beam diagnostics, the liquid hydrogen target, 
the spectrometer/collimator system, and 
detectors. They are described in our previous
publications~\cite{E158prl,theses,scanner}; we discuss them briefly here. 

The high-intensity polarized electron beam, delivered in $\sim 270$~ns
pulses at 
the rate of 120 Hz, passes through a 1.57~m long 
cylindrical cell filled with liquid hydrogen \cite{target}. 
Scattered particles with $4.4<\theta_\mathrm{lab}<7.5$~mr over the
full range of the azimuth
are selected by the magnetic spectrometer \cite{theses},
while the primary beam and 
forward angle photons pass unimpeded to the beam dump.

Sixty meters downstream of the target, 
scattered \Moller\ electrons in the range 13-24~GeV form an
azimuthally-symmetric ring, spatially separated from electrons 
scattered from target protons ($ep$ scattering). 
The charged particle 
flux is intercepted by the primary copper/fused silica fiber sandwich
calorimeter. 
The asymmetry is measured by extracting the fractional difference in the 
integrated calorimeter response for incident right- and left-handed 
beam pulses.

The calorimeter provides both radial and azimuthal segmentation. Four radial
rings are uniformly covered in the azimuth by
10, 20, 20, and 10 photomultipliers, respectively. 
The three inner rings, referred to as the Inner, Middle and Outer 
\Moller\ rings are predominantly sensitive to \Moller\ scattered
electrons.
The outermost or EP ring intercepts the bulk of the $ep$ flux.

The background within the \Moller\ rings is estimated to be $\simeq
8$\%. It is dominated by radiative $ep$ scattering, while
neutral particles and charged pions contribute less than 1\%. 
Quartz detectors placed behind the \Moller\ detector and shielding
record the 
charged pion asymmetry. 
Target density fluctuations and spurious asymmetries are monitored by 
intercepting charged particles at $\theta_{\mathrm {lab}}\approx 1$~mr 
with gas ionization chambers
\cite{lumipaper}.

%% file: analysis.tex
The data sample consists of $2.9\times 10^8$ and $3.7\times 10^8$
pulses at beam energies of $45.0$ and $48.3$ GeV respectively,
collected over three data runs in 2002 and 2003.  
Roughly 60\% of the data were accumulated in the 2003 run,
which featured a novel ``superlattice'' photocathode \cite{superlattice}
with $\approx 90\%$ beam polarization.

Data were collected at 120 Hz, with $\sim 1$ Hz of pulses blanked to
measure baseline signals. Alternate triggers fall into two 60~Hz
fixed-phase ``time slots''. Within these time slots, right-left pulse
pairs are formed for independent asymmetry analyses. The helicity
sequence in each pulse pair was chosen pseudo-randomly. In addition to
the fast helicity flips, the sign of the electron polarization was
passively reversed in two independent ways. First, the state of a
half-wave plate in the laser line was toggled each day, guarding
against helicity-correlated electronics crosstalk. Second, spin
precession in the $24.5^\circ$ bend after acceleration created
opposite helicity orientations at $45$ and $48$ GeV beam energies. 
For each of the 2002 runs the beam energy was changed once, while the
change was made 
roughly every four days during the 2003 run.
Roughly equal statistics were accumulated with 
opposite signs of the measured asymmetry,
suppressing many classes of systematic effects.


We select pulses with beam charge greater than $10^{11}$
electrons and require that the beam crosses each beam position monitor (BPM) 
within 1~mm of its geometric center.
We also require that the beam position and
charge for each pulse be within 6 standard deviations from the running
mean value. Typically, the beam charge per pulse varies
between $(4-6)\times 10^{11}$ electrons with $0.3\%$ pulse-to-pulse
jitter, and the beam position is within $100~\mu$m of each
BPM center, with jitter on the order of $50~\mu$m. 
In order to avoid
helicity-dependent biases, we reject several pulses before and after a
pulse which fails a cut. Other than the demand that
the beam charge asymmetry measured by two independent monitors agree to within
$10^{-3}$, no helicity-dependent cuts are made.

The right-left asymmetry in the integrated detector response for each pulse
pair is computed by normalizing to incident charge and 
then correcting for beam fluctuations.
To first
order, six correlated parameters describe the
beam trajectory: charge, energy, and horizontal and vertical
position and angle. Each parameter is measured by two
independent monitors, such that device resolution and systematic
effects can be studied. 

Two methods are used to calibrate
the detector sensitivity to each beam parameter and remove
beam-induced random and systematic effects from the raw asymmetry. 
One method uses a calibration subset (4\%) of the pulses,
where each beam parameter is modulated periodically
around its average
value by an amount large compared to nominal beam fluctuations.
The other method applies an unbinned least squares linear regression 
to the pulses used for physics analysis.
They yield statistically consistent results to within $3$~ppb.
Final results are obtained with the latter, statistically more powerful
technique. 

%
%
Additional bias to the measured asymmetry may arise from
asymmetries in higher order moments of beam distributions, such as
temporal variations of the beam position or
energy within a 270~ns beam pulse, coupled to the intra-pulse
variation of position or energy asymmetries. Such higher order biases
are small for the Inner and Middle \Moller\ detector rings, but are
observed to be significant for the Outer ring. 

During the physics runs, great care was taken to minimize
residual time structure of the beam position at the target, keeping it
typically within 1 mm. In order to measure the possible
bias due to such effects, six BPMs were instrumented with
23 additional readout channels before the 2003 run. Thus, in addition to 
the average beam parameters for each pulse, BPM signals for charge,
energy, positions, and angles are each digitized in four independent time
slices (three slices for energy). Corrections due to intra-pulse
variation of beam asymmetries are computed by linearly regressing
\Moller\ asymmetries against the beam asymmetries in time slices.  

For the 2003 data, the regression analysis limits the possible
contribution to the detector asymmetry due to the
intra-pulse variations to 3 ppb. Since time slice
data were not available for 2002 data, the Outer \Moller\ ring channels
are only used in 2003 data sample. 
For the 2002 datasets, 
these channels set a limit on the maximum possible bias
in the two innermost \Moller\ rings (containing the bulk of the
statistical weight) of less than 5 ppb. 

After linear regression, the integrated responses of all the 
selected rings are averaged to form the raw asymmetry
$A_{\mathrm{raw}}$.  Near-perfect azimuthal symmetry reduces the
sensitivity to beam fluctuations and right-left beam asymmetries. The
$A_\mathrm{raw}$ pulse-pair distribution has an average RMS of
$215$~ppm for the 2002 data and $185$~ppm for the 2003 data. 
The cumulative beam asymmetry
correction is $-9.7\pm 1.4$~ppb.  A correction due to an azimuthal
modulation of $A_{\mathrm{raw}}$ \cite{transverse} from a small
non-zero transverse component of the beam polarization  is found to be $-3.8\pm
1.5$~ppb.

The average electron beam polarization is
$P_b=0.89\pm 0.04$, measured every few days by a 
polarimeter using \Moller\ scattering of beam electrons
off a magnetized foil. The linearity of the calorimeter response
is determined to be $\epsilon=0.99\pm 0.01$ from special calibration runs.

The physics asymmetry $A_{\rm phys}$ is formed from $A_{\rm raw}$ by
correcting for background
contributions, detector linearity and beam polarization:
\[
A_\mathrm{phys} = \frac{1}{P_b \epsilon} 
\frac{A_\mathrm{raw} - \sum_i \Delta A_i}{1-\sum_i f_i} \, .
\]
Asymmetry corrections $\Delta A_i$ and dilutions $f_i$ for various
background sources are 
listed in \tabref{tab:systematics}. 
%
%
\begin{table}[b]
\caption{Corrections $\Delta A_i$ and dilutions $f_i$ 
to $A_{\rm raw}$ and associated systematic uncertainties.
}
\label{tab:systematics}
\begin{center}
\begin{tabular}{|l|r|c|}
\hline
Source & $\Delta A$ (ppb) & $f$ \\
\hline
Beam (first order)  & $-10\pm 1$& \\
Beam (higher order) & $0\pm 3$& \\
Transverse polarization & $-4\pm 2$ & \\
$e^-+p\rightarrow e^-+p(+\gamma)$ & $-7\pm 1$ & $0.056\pm 0.007$ \\
$e^-(\gamma)+p\rightarrow e^-+X$ & $-22\pm 4$ & $0.009\pm 0.001$ \\
$\gamma+e^-\rightarrow e^-+\gamma$ & $0\pm 1$ & $0.005\pm 0.002$ \\
High energy photons & $3\pm 3$ & $0.004\pm 0.002$\\
Synchrotron photons & $0\pm 1$ & $0.002\pm 0.001$\\
Pions               & $1\pm 1$  & $0.001\pm 0.001$\\
\hline
\end{tabular}
\end{center} 
\end{table}

%% file: results.tex

\begin{figure}
\includegraphics[width=3.3in]{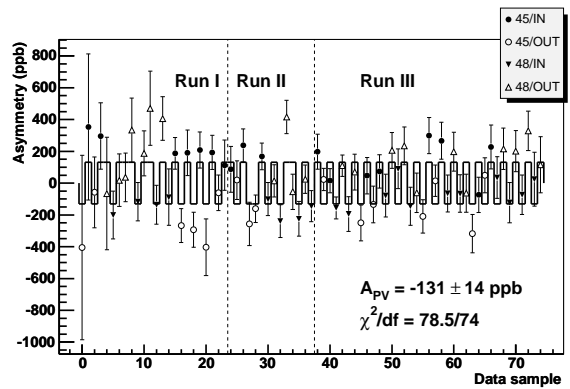}
\caption{\label{fig:slug} $A_{\rm phys}$ for each of 75
data samples. Data collected with half-wave plate inserted(removed) 
at a beam energy of 45(48) GeV are shown
as solid(open) circles(triangles). The solid line represents the grand
average, with the expected modulation of the asymmetry sign 
for each beam energy and half-wave plate state. 
Only statistical uncertainties are shown. 
}
\end{figure}

Figure~\ref{fig:slug} shows $A_\mathrm{phys}$ for all data, 
divided into 75 sequential samples in Runs I, II (2002), and III
(2003). Each $A_\mathrm{phys}$ measurement has  
sign reversals depending on the
beam energy and the state of the half-wave plate.
$A_{PV}$ is obtained by correcting each result by the appropriate
sign. 
The combined result is
\[
A_{PV} = -131\pm 14\ \stat\ \pm 10\ \syst~\mathrm{ppb}\ .
\]

In the context of the Standard Model, we interpret the measurement of
\APV\ 
in terms of the effective weak mixing angle 
$\stwrun(Q)$ \cite{marciano1}:
\[
A_{PV} = -{\mathcal A}(Q^2,y) \rho^{(e;e)}\left[1 - 4\stwrun(Q) + \Delta\right]
\ .
\] 
The average values of the kinematic variables are
$Q^2 = 0.026~\gev^2$ and $y=Q^2/s \simeq 0.6$, where
$s$ is the square of the center-of-mass energy.
\[
{\mathcal A}(Q^2,y) = \frac{G_F Q^2}{\sqrt{2}\pi\alpha(Q)}\frac{1-y}{1+y^4+(1-y)^4}
{\mathcal F}_{\rm QED}
\]
%
is the effective analyzing power, 
$G_F$ and $\alpha(Q)$ are the Fermi and
fine structure constants, 
respectively \cite{PDG2004}, 
$\rho^{(e;e)}$ is the low-energy ratio of
the weak neutral and 
charge current couplings, 
${\mathcal F}_{\mathrm QED}=1.01\pm 0.01$ is a QED radiative
correction factor that includes 
kinematically weighted hard initial and final state radiation 
effects and $y$-dependent contributions from $\gamma\gamma$ and
$\gamma Z$ box 
and vertex diagrams \cite{Zykunov}. 
$\Delta$ contains residual ${\mathcal O}(\alpha)$ electroweak
corrections. 
The effective analyzing power 
${\mathcal A} = 3.25\pm 0.05$~ppm 
is determined from a Monte Carlo simulation 
that accounts for
energy losses in the target and systematic uncertainties in the
spectrometer setup. 

\begin{figure}
\includegraphics[width=3.3in]{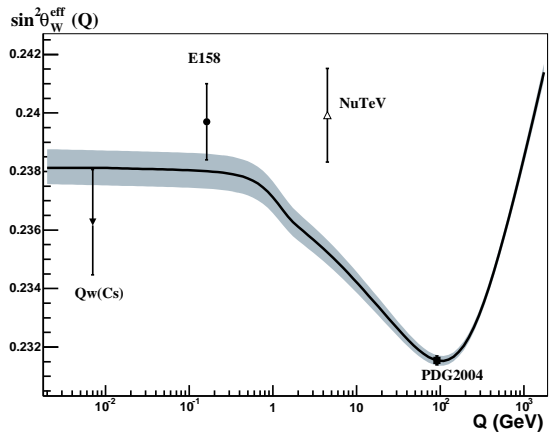}
\caption{\label{fig:running} 
Predicted variation \cite{marciano2} of 
\stwrun\ as a function of 
momentum transfer $Q$ (solid line) and its estimated theoretical uncertainty
(shaded area). Results of prior low energy experiments
\cite{ref:APVcs,PDG2004} (closed triangle, shown at an arbitrarily
higher $Q$) and \cite{ref:nutev} (open triangle) are
overlaid together with the $Z^0$ pole value \cite{PDG2004} (square) and this
measurement (circle).}
\end{figure}

A number of definitions of the low-energy weak mixing angle exist
\cite{marciano2,otherad} and differ in
the way various corrections of order ${\mathcal O}(\alpha)$ are
distributed between terms $\stwrun(Q)$, $\rho^{(e;e)}$, and $\Delta$. 
Here we adopt a definition of the coupling $\stwrun(Q)$
\cite{marciano2} which reproduces the effective leptonic coupling
$\stwrun(M_Z)\equiv \bar{s}_l^2=0.23149\pm 0.00015$ \cite{PDG2004}
at $Z^0$ pole. This 
implies $\rho^{(e;e)}=1.0012\pm 0.0005$ and $\Delta = -0.0007\pm
0.0009$. 
We determine at $Q^2=0.026~\mathrm{GeV}^2$ 
\[
\stwrun(Q) = 0.2397\ \pm 0.0010\ \stat\ \pm 0.0008\ \syst\ .
\]
Our value is consistent with the Standard Model expectation
\cite{marciano1,PDG2004} $\stwrun(Q)=0.2381\pm 0.0006$ and is
$6.2\sigma$ away from $\stwrun(M_Z)$ (Fig.~\ref{fig:running}).
Interpreting our result as a measurement of the electroweak coupling
parameter $\stwmsbar$ yields
\begin{eqnarray*}
\stwmsbar = 0.2330 & \pm & 0.0011\ ({\rm stat.})\\
          \pm\ 0.0009\ ({\rm syst.}) & \pm &  0.0006\ ({\rm theory}).
\end{eqnarray*}
The last uncertainty is from the evolution to $M_Z$.

Our measurement of \APV\ can also be used to set limits on the size of possible
new contributions beyond the Standard Model. Quite generally, 
we set a limit on the scale $\Lambda_\mathrm{LL}$
of a  new left-handed contact interaction characterized by a
term in the  Lagrangian \cite{lambda} 
${\mathcal L}=\pm(4\pi/2\Lambda_\mathrm {LL}^{\pm 2})(\bar{e}_L\gamma_\mu e_L)$.
At 95\%\ C.L. a tree-level calculation yields
$\Lambda_\mathrm {LL}^+\ge 7$ TeV and $\Lambda_\mathrm {LL}^-\ge 16$ TeV, 
for potential positive and negative deviations, respectively. 
As an
example of a specific model, the 95\%\ C.L. on the mass of $Z_\chi$
boson appearing in the grand unified model with SO(10) symmetry
\cite{marciano1,lambda} is 
$M_{Z_\chi}\ge 1.0$~TeV.

%% file: conclusion.tex

In summary, we have reported a new measurement of $A_{PV}$ in
\Moller\ scattering with an accuracy of 17 ppb. 
This leads to a precise determination of 
$\stwrun$ at low momentum transfer. The running of the
weak mixing angle is observed with over $6\sigma$ significance.
The consistency of the result
with the Standard Model prediction provides significant new limits on
TeV scale physics, comparable in sensitivity and complementary to the best
current limits from high energy colliders.

%% file: ack.tex
\begin{acknowledgments}

We thank the SLAC staff for their efforts in helping develop and
operate the E158 apparatus, and especially the Polarized Electron
Source and Accelerator Operations groups.  We would also like to thank
A.~Czarnecki, J.~Erler, W.J.~Marciano, and M.~Ramsey-Musolf for
stimulating discussions, and A.~Ilyichev, J. Suarez, and V.~Zykunov
for providing the radiative correction software.  This work was supported
by Department of Energy contract DE-AC02-76SF00515, and by the
Division of Nuclear Physics at the Department of Energy and the
Nuclear Physics Division of the National Science Foundation in the
United States and the Commissariat \`{a} l'\'{E}nergie Atomique and
the Centre National de la Recherche Scientifique in France.

\end{acknowledgments}

%% file: prl2.bbl
\begin{thebibliography}{99}
\bibitem{zeld}
Ya.B. Zel'dovich, \journal{Sov. Phys. JETP}{94}{262}{1959}.

\bibitem{ref:SM}
S. Weinberg, \PRL{19}{1264}{1967};
A. Salam, in {\em Elementary Particle Theory}, p. 367,
ed. N. Svartholm (Almquist and Wiksells, Stockholm, 1969);
S.L. Glashow, J. Iliopoulos, and L. Maiani, \PR{D2}{1285}{1970}.

\bibitem{ref:running}
E.C.G. Stuckelberg and A. Petermann,
\journal{Helv. Phys. Acta}{24}{317}{1951};
M. Gell-Mann and F. Low, \PR{95}{1300}{1954};
N.N. Bogolyubov and D.V. Shirkov, \journal{Dokl. AN SSSR}{103}{203}{1955}.

\bibitem{marciano1}
A. Czarnecki and W.J. Marciano, \PR{D53}{1066}{1996}.

\bibitem{otherad}
J. Erler, A. Kurylov, and M.J. Ramsey-Musolf, \PR{D68}{016006}{2003};
A. Ferroglia, G. Ossola, and A. Sirlin, 
\journal{Eur. Phys. J.}{C34}{165}{2004};
J. Erler and M.J. Ramsey-Musolf, preprint hep-ph/0409169;
F.J. Petriello, \PR{D68}{033006}{2003}.

\bibitem{ref:APVcs}
S.C. Bennett and C.E. Wieman, \PRL{82}{2484}{1999}.

\bibitem{ref:nutev}
G.P. Zeller \etal, \PRL{88}{091802-1}{2002}.

\bibitem{E158prl}
P. Anthony \etal, \PRL{92}{181602}{2004}.

\bibitem{darman}
E. Derman and W.J. Marciano,
\journal{Ann. Phys.}{121}{147}{1979}.

\bibitem{theses}
Ph.D Theses: 
D. Relyea, B. Humensky (Princeton University, 2003),
I. Younus (Syracuse University, 2003),
K. Bega, P. Mastromarino, G.M. Jones (California Institute of
Technology, 2004),
A. Vacheret (L'Universit\'e Louis Pasteur, 2004),
W. Emam (Syracuse University, 2004).

\bibitem{scanner}
R.S. Hicks \etal, preprint nucl-ex/0504029, submitted to 
Nucl. Insrum. Methods. 

\bibitem{target}
J. Gao \etal, \NIM{A498}{90}{2003}.

\bibitem{lumipaper}
C. Field \etal, \NIM{A531}{569}{2004}.

\bibitem{superlattice}
J.E. Clendenin \etal, \NIM{A536}{308}{2005}. 

\bibitem{transverse}
A.O. Barut and C. Fronsdal, \PR{120}{1871}{1960};
L.L. DeRaad, Jr. and Y.J. Ng, \PR{D11}{1586}{1975};
L.J. Dixon and M. Schreiber, \PR{D69}{113001}{2004}.

\bibitem{PDG2004}
Particle Data Group, S.~Eidelman \etal, \PL{B592}{1}{2004}.

\bibitem{Zykunov}
V. Zykunov, \journal{Yad.\ Phys.}{67}{1366}{2004},
\journal{Phys.\ Atom.\ Nucl.}{67}{1342}{2004}.

\bibitem{marciano2}
A. Czarnecki and W.J. Marciano, \journal{Int. J. Mod. Phy.}{A15}{2365}{2000}.

\bibitem{lambda}
E.J.~Eichten, K.D.~Lane, and M.E.~Peskin, \PRL{50}{811}{1983};
M.J. Ramsey-Musolf, \PR{C60}{015501}{1999}.


\end{thebibliography}
